\documentclass[runningheads]{llncs}

\usepackage[T1]{fontenc}

\usepackage{graphicx}
\usepackage{color}

\usepackage{enumitem}
\usepackage{algorithm}
\usepackage{float}
\usepackage{hyperref}

\usepackage{booktabs}
\usepackage{multirow}

\usepackage{url}

\usepackage{orcidlink}

\usepackage{subcaption}

\begin{document}

\title{Selfish Mining in Multi-Attacker Scenarios: An Empirical Evaluation of Nakamoto, Fruitchain, and Strongchain}

\titlerunning{Selfish Mining in Multi-Attacker Scenarios}

\author{Martin Perešíni\inst{1}\orcidlink{0000-0002-2875-9567} \and
Tomáš Hladký\inst{2} \and
Jakub Kubík\inst{2} \and
Ivan Homoliak\inst{1}\orcidlink{0000-0002-0790-0875}
}

\authorrunning{Perešíni et al.}

\institute{Brno University of Technology, Božetěchova 2, Brno, Czech Republic, 612 00\\ \email{\{iperesini, homoliak\}@fit.vut.cz} \and \email{\{xhladk15, xkubik32\}@stud.fit.vut.cz}}

\maketitle 

\begin{abstract}
The aim of this work is to enhance blockchain security by deepening the understanding of selfish mining attacks in various consensus protocols, especially the ones that have the potential to mitigate selfish mining.
Previous research was mainly focused on a particular protocol with a single selfish miner, while only limited studies have been conducted on two or more attackers.
To address this gap, we proposed a stochastic simulation framework that enables analysis of selfish mining with multiple attackers across various consensus protocols.
We created the model of Proof-of-Work (PoW) Nakamoto consensus (serving as the baseline) as well as models of two additional consensus protocols designed to mitigate selfish mining: Fruitchain and Strongchain.
Using our framework, thresholds reported in the literature were verified, and several novel thresholds were discovered for 2 and more attackers.
We made the source code of our framework available, enabling researchers to evaluate any newly added protocol with one or more selfish miners and cross-compare it with already modeled protocols.
\end{abstract}

\section{Introduction}

Proof-of-Work (PoW) blockchains such as Bitcoin rely on Nakamoto consensus~\cite{nakamoto2009bitcoin} to agree on a chain of blocks.
However, in 2013, Eyal and Sirer~\cite{selfish_1} showed that a miner who only controls 33\% of the hash rate could earn more than their fair share through selfish mining (SM), challenging the original assumption that the majority $>$50\% power is needed for such an attack.
The selfish miner strategically withholds and releases blocks that override the honest chain, thus wasting the work of honest miners.
Since its discovery, extensive research has explored the impact and potential mitigations of selfish mining~\cite{gal2023selfish},~\cite{insightfulmining}.
The follow-up works~\cite{optimal_sm},~\cite{selfish_2_b},~\cite{selfish_m} successively tightened the single-attacker profitability threshold to 25\%, and in the settings of two attackers to 21\% per attacker (up to ten attackers).
Another variant is stubborn mining~\cite{stubborn}, in which attackers persist in mining their private chains even when they fall behind the honest chain, potentially increasing their profits.

Parallel to the analytic results, researchers proposed fork-choice tweaks (e.g., uniform tie breaking) or entirely new PoW schemes: Subchain~\cite{subchain}, Fruitchain~\cite{fruitchain}, and Strongchain~\cite{strongchain}, which are aimed at mitigating SM by increasing the SM profitability threshold, thus making SM attacks much more expensive and less feasible.
Nevertheless, the literature still lacks a comparative multi-attacker evaluation of these defenses under identical conditions.
Our work aims to address this gap by providing an empirical evaluation of selfish mining strategies across the mentioned consensus protocols.
We introduce a unified simulation framework capable of comparing various consensus protocols under selfish mining attacks.
Our study explores scenarios with one or more selfish miners and investigates the impact of the attacker's connectivity (i.e.,  propagation factor ($\gamma$)) on profitability thresholds of selfish mining.

\noindent
\paragraph{\textbf{Contributions.}}
Our main contributions are as follows:
\begin{enumerate}
    \item We created a unified simulation framework capable of modeling various consensus protocols under selfish mining attacks.

    \item We implemented and analyzed designs to mitigate selfish mining, particularly Fruitchain~\cite{fruitchain} and Strongchain~\cite{strongchain} within this framework, allowing direct comparisons.

    \item We performed extensive simulations exploring selfish mining scenarios with one or more attackers, revealing new profitability thresholds under various values of the propagation factor parameter (i.e., $\gamma$).

    \item We cross-compared the investigated consensus protocols in terms of selfish mining mitigation and revealed their relative order of protection.
\end{enumerate}

\section{Background on Selfish Mining}

In a selfish mining attack, the attacker has a consensus power below $50\%$ of the total mining power of the blockchain. The primary aim of a selfish miner is to gain a higher reward relative to their consensus power~\cite{selfish_1}. 

The selfish miner selectively reveals his mined blocks to override and invalidate the honest miners' work. He keeps his mined blocks private, creating a private chain. Since the~selfish miner controls a relatively small portion of the total mining power, his private branch can only stay ahead for a short time. During this period, honest miners continue to mine on the shorter public branch. 
A selfish miner conveniently reveals blocks from his private branch, overriding the shorter public branch, which causes honest miners to abandon the public branch and switch to the recently revealed attacker's chain. This wastes the work done on the public branch by honest miners and increases the fraction of selfish miners' blocks incorporated into the blockchain, allowing them to collect relatively more revenues than invested mining power~\cite{selfish_1}.

In a selfish mining attack, the attacker performs 4 actions \cite{selfish_1}:
\begin{itemize}
    \item[\textbullet] \textbf{Override} -- if the honest main chain is just one block shorter than the attacker's private chain, the attacker publishes his private chain to override the main chain and obtain rewards.
    
    \item[\textbullet] \textbf{Adopt} -- the attacker accepts the honest main chain and abandons his private chain when the main chain becomes longer than the attacker's private chain.
    
    \item[\textbullet] \textbf{Match} -- if the main chain is as long as the attacker's, the attacker publishes his private chain to compete with the honest main chain.
    
    \item[\textbullet] \textbf{Wait} -- if the attacker's chain is more than one block longer than the honest main chain, the attacker continues to mine on his private chain.
\end{itemize}

\subsection{Notation}
\label{notation}

\paragraph{\textbf{Mining Power.}}
The blockchain system consists of the set of miners $M = \{m_i\}$, where $i \in  \langle 1, \dots, n \rangle$. 
Each miner $i$ has a~mining power of $\alpha_i$, and the total mining power of all miners is equal to $1$; i.e., $\sum_{i=1,\ldots,n}{\alpha_i} = 1$. 

\paragraph{\textbf{Propagation Factor.}}
During the selfish mining attack, the attacker releases his private chain in the \textbf{match action} (i.e., equal length of the honest and the malicious chains) and continues to mine on it. Honest miners decide which branch to continue mining on, usually by selecting the branch they receive first. 
The proportion of honest miners who choose to mine on the chain revealed by the attacker is represented by $\gamma \in \langle 0,1 \rangle$, called \textbf{a~propagation factor}.
If $\gamma = 0$, the honest miner always wins the race of the block propagation to the network. 
Contrary, when $\gamma = 1$, the selfish miner always wins the race of the block propagation to the network.
If $\gamma = 0.5$, both honest and selfish miners win the race of block propagation with equal chances.
The propagation factor can also be abstracted as the connection quality of the attacker vs. the honest miners; therefore, the higher $\gamma$, the better connectivity the attacker has vs. the honest miners.

\subsection{Selfish Mining with a Single Attacker}
\label{one_attacker}
\begin{figure}[t]
    \centering
    \begin{minipage}[t]{0.48\textwidth}
        \centering
        \includegraphics[width=\linewidth]{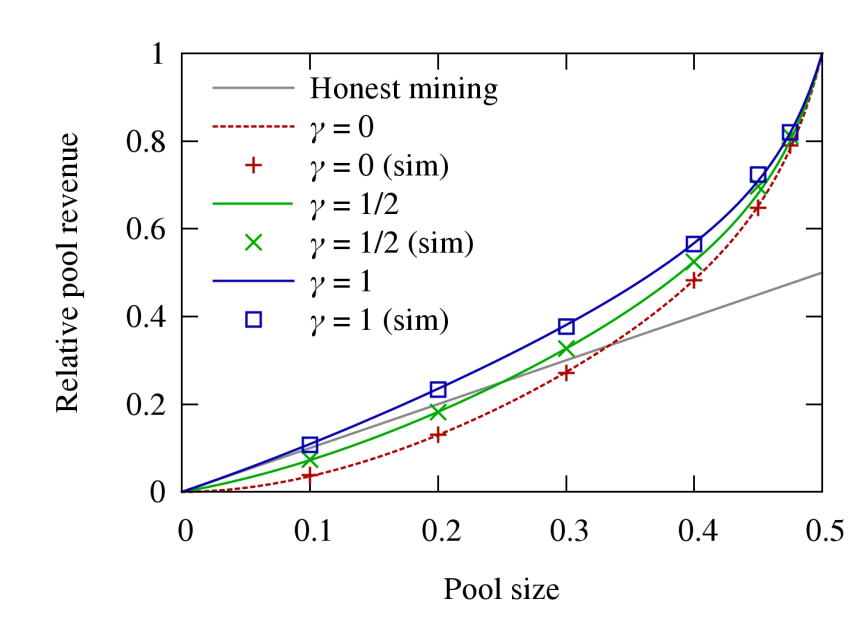}
        \caption{Selfish mining with a single attacker on Nakamoto's consensus~\cite{selfish_1}.}
        \label{fig:selfish_m}
    \end{minipage}
    \hfill
    \begin{minipage}[t]{0.45\textwidth}
        \centering
        \includegraphics[width=\linewidth]{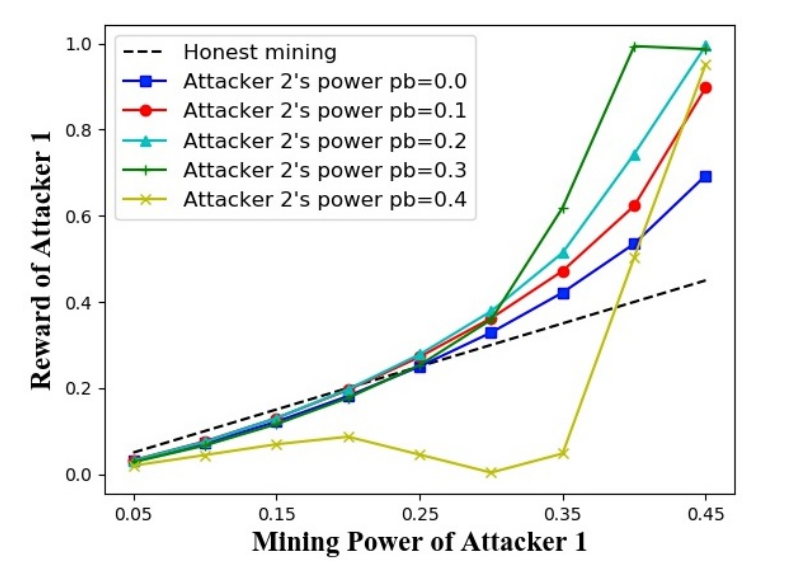}
        \caption{Selfish mining with 2 attackers~\cite{selfish_2_b}: rewards of the attacker 1 based on his mining power vs. the mining power of attacker 2.}
        \label{fig:two_rew_one}
    \end{minipage}
    \vspace{-1em}
\end{figure}

Eyal and Sirer~\cite{selfish_1} performed modeling by Markov chains and executed simulations of selfish mining by a single attacker on Nakamoto's consensus~\cite{nakamoto2009bitcoin} (see \autoref{fig:selfish_m}).
The $x$-$axis$ of \autoref{fig:selfish_m} displays various mining powers $\alpha$ of an attacker, while the $y$-$axis$ represents their relative revenues.
The figure compares various values of $\gamma$ between selfish miners and honest miners. 
It shows that selfish mining becomes profitable for a poorly connected attacker ($\gamma=0$) at the mining power $\sim$33\%, while for an equally connected attacker ($\gamma=0.5$) the attack becomes profitable at $\sim$25\%.

\subsection{Selfish Mining with Two Attackers}
\label{two_attackers}

\begin{figure*}[b]
    \vspace{-0.5cm}
	\centering
	\includegraphics[width=\textwidth]{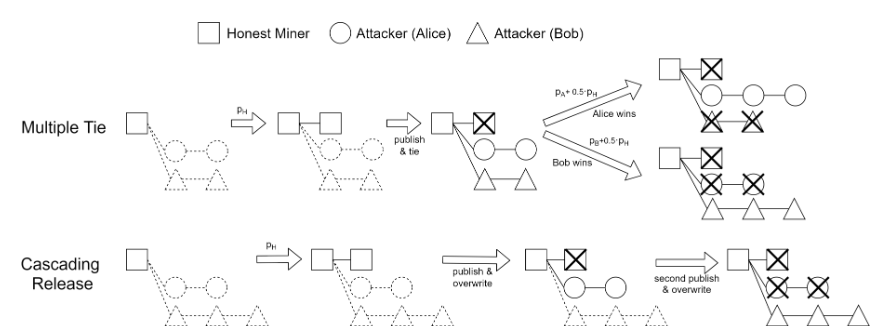}
	\caption{Selfish mining with 2+ independent attackers yields new scenarios: (1) multiple ties and (2) cascading release~\cite{selfish_2_b}.}
	\label{fig:two_self_cases}
    \vspace{-1em}
\end{figure*}

In the case of 2 or more selfish miners (on Nakamoto's consensus), multiple private branches can co-exist simultaneously since each attacker privately mines on their chain.
Notation for multiple attackers.
Let $p_A$ and $p_B$ denote the mining powers of the attackers Alice and Bob, respectively, and $p_H=1-p_A-p_B$ the total honest mining power.
When both attackers publish competing private branches of equal length ("multiple tie"), the probability that Alice's branch wins is $p_A+\frac{1}{2}p_H$, and Bob's is $p_B+\frac{1}{2}p_H$.
Such a setting enables 2 new scenarios in contrast to selfish mining by a single attacker~\cite{selfish_2_b} (see \autoref{fig:two_self_cases}):
\begin{itemize}
    \item[\textbullet] In a \textbf{multiple tie}, multiple branches of the same length are published by the attackers.
    For example, attackers Alice and Bob have a private chain with two blocks each. Honest miners will randomly select one of the two branches.
    Alice's chain wins with a probability of $p_A + 0.5 * p_H$, and Bob's chain wins with a probability of $p_B + 0.5 * p_H$.
	
    \item[\textbullet] In a \textbf{cascading release}, one attacker overrides the chain of another attacker.
    In the example, Alice and Bob each hold a private branch with lengths of two and three, respectively. When an honest miner mines a new block, Alice releases her branch, overwriting the public chain.
    Then Bob discovers a public chain one block shorter than his private branch, so he releases his private branch, overwriting Alice's branch. 
\end{itemize}

\medskip
\noindent
Zhang et al.~\cite{selfish_2_b} simulated 2 selfish miners on Nakamoto's consensus and their results are depicted in \autoref{fig:two_rew_one}, which displays the rewards of attacker 1 based on his mining power and the mining power of attacker 2. 
We can see that the blue series represents a selfish mining attack with one attacker (and $\gamma=0.5$) since the mining power of attacker 2 is $0\%$. 
Also, as the mining power of the attacker 2 increases (up to 40\%), the reward for the attacker 1 also increases as long as the mining power of the attacker 1 is greater than or equal to the mining power of the attacker 2. 
However, if attacker 2 has $40$+\% (yellow series) and attacker 1 has lower mining power than attacker 2, attacker 1 loses significantly. 
This is because a dominant miner's chain is likely to be always longer and win more often than other chains.
In further simulations, Zhang et al.~\cite{selfish_2_b} demonstrate that the minimum $\alpha$ for which both attackers can profit in this scenario is equal to 21\% (i.e., $42\%$ in total).

\subsection{Selfish Mining with Multiple Attackers}
\label{more_attackers}

Zhang et al.~\cite{selfish_m} conducted multiple simulations on Nakamoto's consensus with three to ten independent attackers.
The authors conclude that the beneficial threshold for selfish mining decreases as the number of attackers increases. 
According to their simulation, seven attackers with at least $12\%$ of the total mining power are the maximum number of successful attackers.
In the case of five attackers, they must hold at least $15\%$ of the total mining power.
Beyond this and former works we extend, multi-player selfish-mining dynamics have also been studied by Marmolejo-Cossio et al.~\cite{marmolejo} and Kwon et al.~\cite{kwon}, who analyze competitive equilibria and strategic interactions among multiple selfish miners.
Our empirical framework complements these by providing cross-protocol thresholds under identical conditions.

\subsection{Threat Model \& Assumptions}\label{threat_model}
Recall that $M$ is the set of all miners with mining powers $\{m_i\}^n_{i=1}$, where $\sum m_i = 1$.
A subset $\mathcal{A}$ of $M$ ($where\ |A| \geq 1$) represents selfish miners and the rest represents honest miners.
We consider rational non-colluding selfish miners competing against honest miners in PoW-based protocols.
Selfish miners follow the Eyal-Sirer's~\cite{selfish_1} selfish mining strategy (i.e., override/match/wait/adopt), adapted according to the protocol fork choice rule.
Also, we assume that honest miners always follow the protocol.

\paragraph{\textbf{Network model.}}
We abstract the network propagation of the attacker by the standard $\gamma$ parameter: when branches tie, a $\gamma$ fraction of the honest mining power mines on the attacker’s branch ($\gamma$=0 honest-favored; $\gamma$=0.5 equal; $\gamma$=1 attacker-favored).
For multi-attacker experiments, we set $\gamma$=0.5 to avoid privileging any single attacker.
Accidental forks and heterogeneous delays are abstracted, since their impact is covered by $\gamma$.
Transaction fees are excluded  across all protocols to isolate side effects of divergent conditions (see also \autoref{conclusion})

\paragraph{\textbf{Attackers' coordination.}}
Unless stated otherwise, attackers act independently and do not collude or share private branches.
Each selfish miner maximizes its own expected relative revenue.

\paragraph{\textbf{Protocol specifics.}}
We implement each protocol’s fork resolution and artifacts (i.e., blocks, fruits, weak headers) according to their descriptions.

\subsection{Robustness Metric: Profitability Threshold}
Let $R_i(\alpha_i)$ denote the relative revenue function of the miner $i$ 
when the miner $i$ controls the mining power $\alpha_i$ (\autoref{fig:fruit_longest}).
We define the profitability threshold $\alpha_i^{\backslash^\ast}$ for the miner $i$ as the smallest $\alpha_i$ such that $R_i(\alpha_i) \geq \alpha_i$.
Intuitively, $\alpha_i^{\backslash^\ast}$ is the minimum power at which selfish mining becomes more profitable than honest mining.
In simulations, we experiment with $\alpha_i$ and estimate $R_i(\alpha_i)$ as the average over five runs of 100{,}000 rounds each (per configuration), then report the smallest $\alpha_i$ where the mean exceeds the fair-share.
When two adjacent grid points straddle equality, we linearly interpolate and confirm that the 95\% bootstrap confidence interval includes the crossing.
When multiple attackers are present, we report per-attacker thresholds under symmetric allocations unless otherwise noted, e.g., $k=2$ attackers each with $\alpha$ (see \autoref{tab:comparison-thresholds}).

\section{Consensus Protocols Mitigating Selfish Mining}\label{sec:att_res_protos}
We briefly review some consensus protocols that are discussed in the literature as candidates to mitigate the impact of selfish mining.
A key idea in this area involves ``inclusive`` protocols~\cite{inclusive} that incorporate and reward work from otherwise orphaned blocks, thereby reducing the attacker's ability to waste honest miners' efforts.
However, these protocols are built on the assumption that no honest miners will willingly put conflicting transactions into their block to potentially increase their profits, which was refuted in the related work~\cite{perevsini2023incentive}.
Beyond protocols that add new artifacts to the chain (e.g., Strongchain~\cite{strongchain} and Fruitchain~\cite{fruitchain}), other approaches modify the block structure and the reward scheme itself to disincentivize withholding blocks~\cite{eyal,subchain}.
For example, Bitcoin-NG~\cite{eyal} proposes a leader-election mechanism with subsequent microblocks, which reduces the per-block reward advantage a selfish miner can gain, thus altering the attack's fundamental economic incentives.
This protocol seems sound, but it has a known problem with censoring a leader, which brings another attacking vector instead of selfish mining.
The Fruitchain and Strongchain protocols, which we evaluate, approach this problem in a different way.
	
\subsection{\textbf{Fruitchain}} 

\begin{figure}[t]
    \centering
    \includegraphics[width=0.7\textwidth]{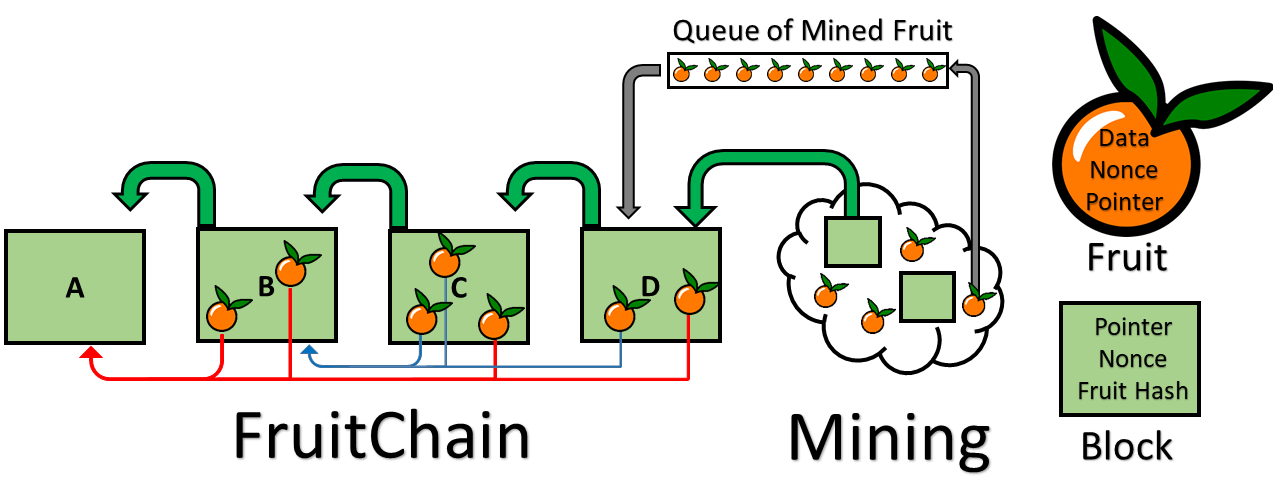}
     \vspace{-0.3cm}
    \caption{A Fruitchain's fragment with fruits and blocks~\cite{fruitchain}.}
    \label{fig:Fruitchain_struct}
    \vspace{-0.5cm}
\end{figure}

The primary objective of the Fruitchain~\cite{fruitchain} consensus protocol is to \textbf{enhance} security, particularly in terms of \textbf{resistance to selfish mining}. The most significant architectural change is that the transactional records on Fruitchain are stored in \textbf{fruits}, which are then stored in \textbf{blocks} (see \autoref{fig:Fruitchain_struct}). 
Individual blocks are used to build the blockchain. 
Creating fruits and blocks requires different levels of PoW difficulty. Additionally, fruits must ``hang'' from a block that is already part of the blockchain. 
\textbf{Hanging fruit} contains a reference to an already existing block on the blockchain within a maximum distance. The maximum distance is determined by a constant value of the protocol.
	
In each round, the miner searches for fruits and blocks simultaneously by invoking a single hash function, a technique called the \textbf{2-for-1 trick}. 
The prefix of the generated hash is the fruit mining hash, while the suffix determines the block mining hash. 
This means that each newly mined block or fruit contains redundant and unused data.
Besides, the authors do not discuss the possibility and handling of duplicate transactions in different fruits since miners are selecting transactions to be included in fruits based on their fees, while fruits are mined more frequently than blocks, which increases the probability that two or more different fruits contain the overlapping set of transactions.

\textbf{Mining Difficulty.}
The Fruitchain protocol is parameterized with two mining difficulties. The first one (which is of higher difficulty) is for mining blocks, and the second one (which is of lower difficulty) is for mining fruits. 
	
\textbf{Rewarding Scheme.}
The authors of Fruitchain argue that the Fruitchain rewarding scheme is more fair than Nakamoto's one because it rewards fruits and blocks. 
However, they do not define particular values nor how fruits and blocks are rewarded.
They also do not define how many transactions are in individual fruits and how are transaction fees attributed.
	
\textbf{Resistance to Selfish Mining.}
The authors of Fruitchain did not assess its resistance to selfish mining. 
Nevertheless, Zhang and Preneel~\cite{sm_improvements} evaluated its resistance to selfish mining with a single attacker. 
The results show that Fruitchain with $\gamma=0$ (i.e., the most favorable setting to honest miners) performs worse than the Nakamoto consensus for the same $\gamma$, which means that the selfish miner in Fruitchain earns higher rewards than in the Nakamoto consensus.
In Fruitchain with $\gamma=1$, the~attacker receives less reward than in the~Nakamoto consensus, so Fruitchain performs better than Nakamoto for $\gamma=1$.

\subsection{\textbf{Strongchain}}
The main objective of the Strongchain~\cite{strongchain} protocol is to improve the~security of the Nakamoto's consensus. 
A~key difference between Strongchain and Nakamoto's consensus is that Strongchain also ''includes`` headers with hash values that do not meet the strong target, but are strong enough to demonstrate significant PoW spent.
Using two mining difficulty levels, it generates two distinct header types: weak and strong. 
Weak headers contribute to the security of the protocol, whereas transactions are stored in strong blocks associated with strong headers.
The blocks contain transactions, weak headers, and a strong header that authenticates the weak headers and transactions. 
The fields in the strong and weak headers are inherited from the Bitcoin headers with an additional new field -- the coinbase entry (the miner's address to receive a reward for a particular strong or weak block). 
Weak headers do not authenticate transactions. They are stored and exchanged without the corresponding transactions. 
A~weak header primarily reflects and contributes to the accumulation of mining power concentrated in a specific branch of the blockchain -- and thus increases the resistance to selfish mining.

\textbf{Mining Difficulty.}
Strong headers, combined with weak headers, are used for chain strength evaluation. 
The~mining difficulty of weak and strong headers is adjusted in every $2016$ blocks in a~predefined ratio. 
Strongchain does not use reward halving.

\textbf{Rewarding Scheme.}
In Strongchain, both weak and strong header miners are rewarded. 
The miner who discovers the strong header receives the full reward $R$, while the weak header miner receives a fraction of $R$, calculated as $\kappa * c R * \frac{T_s}{T_w}$. 
$\kappa$ influences the relative impact of weak header rewards and $c$ is a scaling constant, $R$ represents the reward for strong headers, and $\frac{T_s}{T_w}$ is the difficulty factor of strong headers compared to the difficulty of weak headers. 
Transaction fees belong to the strong header miner.

\textbf{Resistance to Selfish Mining.}
The ratio between the strong header and weak header difficulty influences the number of weak headers that are mined in each round, which in turn influences the threshold of resistance to selfish mining.
For example,  if the~ratio between strong and weak headers difficulty is $1024$, the threshold is around $43\%$, while the threshold is $40\%$ if this ratio is $8$~\cite{strongchain}.

\section{Evaluation}\label{sec:eval}
We proposed a simulation framework capable of simulating and cross-comparing all the simulated consensus protocols. 
In the following, we first describe our simulation framework and then describe the experiments and their results.

\subsection{Simulation Framework}
The proposed simulation framework consists of four primary entities: \textbf{simulation manager}, \textbf{honest miners}, \textbf{selfish miners}, and the \textbf{blockchain}. 
The framework features a centralized entity known as the simulation manager, which replaces the distributed blockchain topology. 
The~PoW mining function is replaced by a uniform random function that selects the leader of a new round based on their mining power. 
The chosen leader (either malicious or honest) creates a new block or other artifacts of the protocols (such as fruits~\cite{fruitchain}, weak headers~\cite{strongchain}).
The leader performs actions depending on whether he is malicious or honest.
Once the leader has taken action, other attackers (selfish miners) can respond if the action is publicly disclosed. 
Afterward, the mining round closes and the process of selecting a new leader repeats until the desired number of rounds is reached.

The simulation framework includes a rewarding scheme for blocks (and/or other artifacts); however, transaction fees are not modeled since they contribute to the overall profit of either malicious or honest miner in the same way within all the investigated consensus protocols.
The framework also implements fork-resolution rules specific to individual consensus protocols. 
Forks can occur only due to selfish mining behavior, while accidental forks\footnote{Contributing to the stale blocks/artifacts.} are not considered since they affect both malicious and honest miners in the same way. 
For simplicity, we abstract from the variance in the propagation and validation times of individual blocks (i.e., the network propagation delay and the validation of blocks are constant). 
Instead, we approximate a difference in network propagation delay between honest miners and selfish miners by the $\gamma$ parameter, which determines the probability that honest miners will accept the attacker's block more likely than the honest miners' block when a match action occurs (see \autoref{threat_model}).

\textbf{Implementation Details.}
We developed the framework using Python 3.8 and TypeScript, and its source code is available on GitHub.\footnote{Repo: \url{https://github.com/Tem12/fruitchain-sim} and \url {https://github.com/Jakub-Kubik/smasf}.} 
The correctness of the implementation was verified by comparing its simulation results with the literature related to Nakamoto's consensus and known configurations of investigated protocols (i.e., for a single selfish miner and particular $\gamma$ if available).

\subsection{Methodology of Experiments}
\label{methodology}
For each investigated consensus protocol, we created its model in our framework and performed several simulation experiments (i.e., runs) with each model.
The experiments contained various parameters such as a different number of selfish miners, the ratio of their mining power, and the parameter $\gamma$ (if applicable). 
Therefore, more than a thousand experiments were often performed for each modeled consensus protocol.
Moreover, we repeated each experiment 5 times per configuration and averaged the results. 
Finally, each simulation experiment involved $100,000$ rounds of consensus protocol,\footnote{Each round produces an artifact -- either a block or a fruit or a weak header or a strong block, depending on the protocol.} allowing a sufficient number of selfish mining states and situations (with multiple selfish miners) to occur during each simulation run.
For each configuration, we identify the smallest $\alpha$ where the mean relative revenue curve intersects the fair-share line.
We additionally verify that the mean exceeds fair-share for the value of $\alpha$ across runs and a 95\% bootstrap confidence interval includes the crossing.
This protects against declaring thresholds based on noisy single-run outliers.

\subsection{Nakamoto Consensus}
To verify the results of related work, we experimented on the Nakamoto protocol with one attacker and $\gamma= \{0, 0.5, 1\}$ as well as for 2 to 7 attackers and $\gamma=0.5$.
In all multi-attacker experiments, we use $\gamma=0.5$ to avoid biasing the fork race toward any single attacker.
The rewards were calculated as the ratio of the blocks mined and included in the blockchain by the miner vs. all the blocks on the blockchain.
The results are presented in \autoref{tab:comparison-thresholds}.
Our findings align with existing research for an attacker with $\gamma = 0.5$ and $\gamma =~1$.
For five and seven attackers, the results deviate by $1\%$.
These results confirm the~validity of our simulation framework with respect to previous research~\cite{selfish_2_b,selfish_m}.

\begin{table*}[t]
\centering
\scriptsize
\begin{tabular}{@{}ccll l lll l lll@{}}
\toprule

\multicolumn{1}{l}{\multirow{2}{*}{\begin{tabular}[c]{@{}l@{}}\textbf{Selfish} \\ \textbf{Miners}\end{tabular}}} & \multicolumn{1}{l}{\multirow{2}{*}{$\gamma$}} & \multicolumn{2}{c}{\textbf{Nakamoto}} &  & \multicolumn{3}{c}{\textbf{Strongchain}} &  & \multicolumn{3}{c}{\textbf{Fruitchain}} \\
\cmidrule(lr){3-4} \cmidrule(lr){6-8} \cmidrule(l){10-12}
\multicolumn{1}{l}{} & \multicolumn{1}{l}{} & \multicolumn{1}{c}{\textbf{Liter.}} & \multicolumn{1}{c}{\textbf{Sim.}} & & \multicolumn{1}{c}{\textbf{Liter.}} & \multicolumn{1}{c}{\textbf{Sim.}} & $\Delta$ & & \multicolumn{1}{c}{\textbf{Liter.}} & \multicolumn{1}{c}{\textbf{Sim.}} & $\Delta$ \\
\midrule

1 & 0 & 33 & 33 & & 45 & 46 & +13 & & N/A & 38 & +5 \\
1 & 0.5 & 25 & 25 & & N/A & - & - & & N/A & 38 & +13 \\
1 & 1 & 1 & 1 & & N/A & - & - & & N/A & 38 & +37 \\
\midrule
2 & 0.5 & 21 & 21 & & N/A & 32 & +11 & & N/A & 38 & +5 \\
3 & 0.5 & 19 & 19 & & N/A & 24 & +5 & & N/A & 25 & +6 \\
5 & 0.5 & 15 & 14 & & N/A & 17 & +3 & & N/A & 17 & +3 \\
7 & 0.5 & 12 & 11 & & N/A & 13 & +2 & & N/A & 13 & +2 \\
\bottomrule
\end{tabular}
\vspace{1em}
\caption{A comparison of selfish mining thresholds [\%] for investigated protocols in contrast to Nakamoto consensus (i.e., denoted by $\Delta$). Note that all the results along N/A in the literature were obtained for the first time in this work.}
\label{tab:comparison-thresholds}
\end{table*}

\subsection{Strongchain}
First, we verified the profitability threshold for a single selfish miner from the literature~\cite{strongchain}. 
Our obtained threshold was by $1\%$ of mining power higher in contrast to the results reported in~\cite{strongchain}.
The rewards were set to $1$ for the strong block and $1$ divided by the weak-to-strong header ratio for weak headers. 
This ratio indicates how many weak headers are mined on average before mining a strong block. 
We set this ratio to 10 for all experiments.
In the original experiments~\cite{strongchain}, the match action of selfish mining always favored honest miners; therefore, we assumed $\gamma = 0$ for all experiments.
The profitability thresholds are shown in \autoref{tab:comparison-thresholds}.
For two simultaneous attackers, the threshold was found to be $32\%$ of the total mining power per attacker.

In our experiments, we explored various scenarios, including single and multiple attackers.
For the single attacker case, our results showed a slightly higher threshold (i.e., by 1$\%$) than reported in the original Strongchain paper~\cite{strongchain}.
This small difference could be caused by differences in the simulation parameters or the stochastic nature of the simulations.
For multiple attackers, we found that Strongchain maintained a relatively high threshold compared to Nakamoto consensus. 
With two simultaneous attackers, the threshold was 32\% of the total mining power per attacker, which is significantly higher than the Nakamoto consensus.
We also investigated scenarios with more than two attackers, finding that Strongchain continued to provide strong resistance against selfish mining even as the number of attackers increased.
This suggests that Strongchain's design, particularly its use of weak headers to contribute to chain strength, provides robust protection against selfish mining attacks.

\begin{figure}[t]
    \centering
    \includegraphics[width=0.48\linewidth]{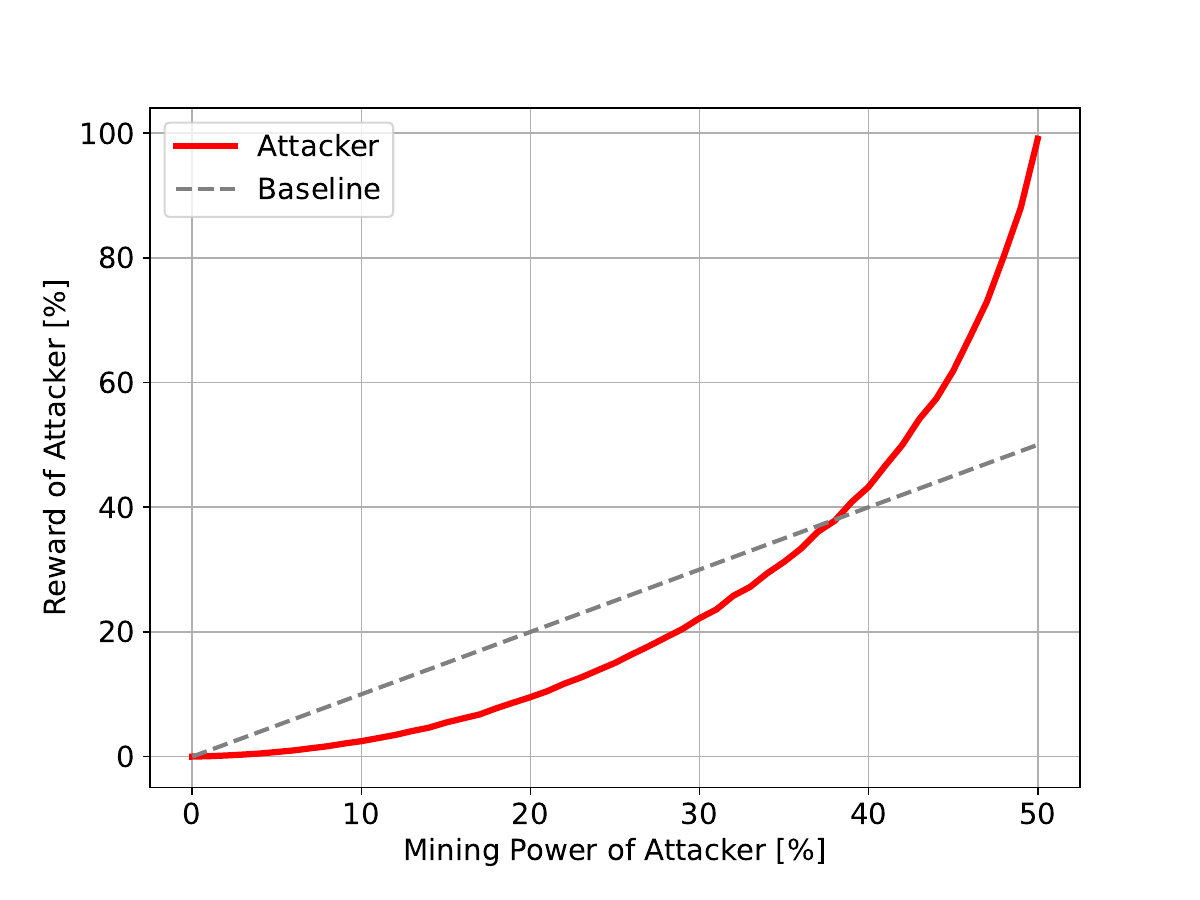}
    \caption{Fruitchain: longest chain rule; block reward = 9$\times$ fruit reward.}
    \label{fig:fruit_longest}
\end{figure}

\subsection{Fruitchain}
Our simulation model incorporated the Fruitchain's unique structure of fruits and blocks.
We set up the simulation to reflect the two-tier mining process, where miners simultaneously mine fruits and blocks.
The simulator configuration contains parameters where the selfish node's mining power was gradually increased and a constant fruit-to-block mining ratio, which is used for weighted random selection, whether fruit or block is being mined.
Based on the ratio, the reward for the fruit is the block mining probability to the fruit mining probability. 
We experimented with different reward ratios for blocks and fruits. 
In particular, we used 50\% and 9.09\%\footnote{9.09 is a consequence of the ratio 10:1 mentioned above (i.e., 1 / 11 * 10).} for block rewards.

In our experiments, we explored various scenarios.
For the single attacker, we found that the profitability threshold is around 38-39\%, which is an improvement over the Nakamoto consensus.
This result was achieved in a configuration where the longest chain rule was used, see \autoref{fig:fruit_longest}.

\subsubsection{Gamma Parameter.}
The parameter $\gamma$ is impactless, even when we choose $\gamma = 0$ or $\gamma = 1$, the variation is only around 1\% between the results.
This is caused by the fact that the override action of selfish mining very rarely happens due to the fine-grained computation of branch difficulty that also considers fruits.\footnote{We note that this phenomenon also occurs in the case of Strongchain.}

\begin{figure*}[t]
	\centering
    \hspace*{-1.35cm}
	\includegraphics[width=1.25\linewidth]{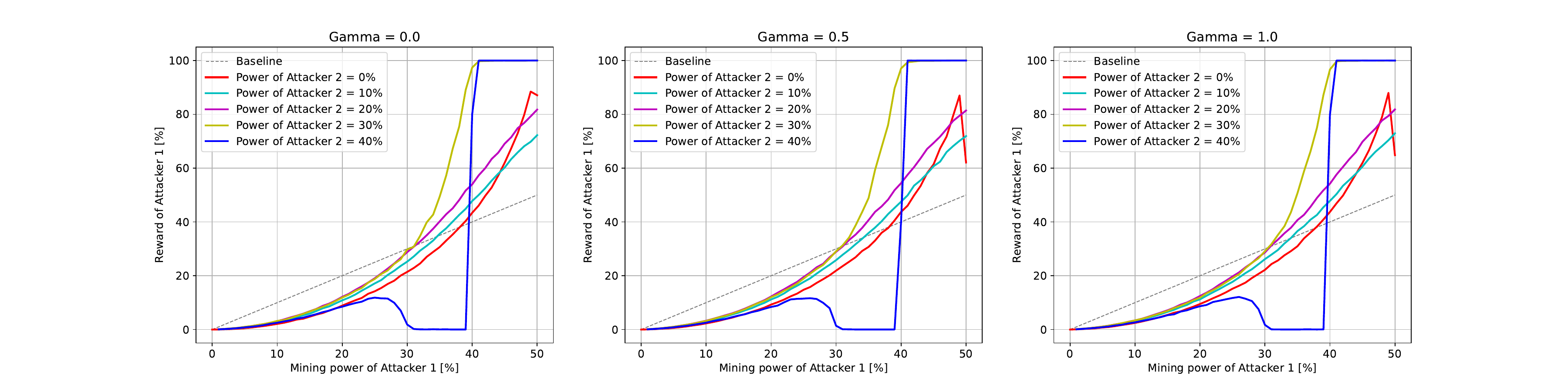}
    \caption{Fruitchain, longest-chain rule. Relative revenue of Attacker 1 versus its mining power $\alpha_1$ under $\gamma \in \{0,0.5,1\}$.
    Each line fixes the other attacker mining power $\alpha_2 \in \{0,10,20,30,40\}$\%.
    The profitability threshold is the intersection with the diagonal baseline $R(\alpha_1) = \alpha_1$.
    As $\alpha_2$ increases, this threshold decreases, lowering Attacker 1's threshold.}
	\label{fig:fruit_longest_two}
\end{figure*}

\begin{figure*}[t]
    \centering
    \begin{subfigure}[b]{0.48\textwidth}
        \centering
        \includegraphics[width=\textwidth]{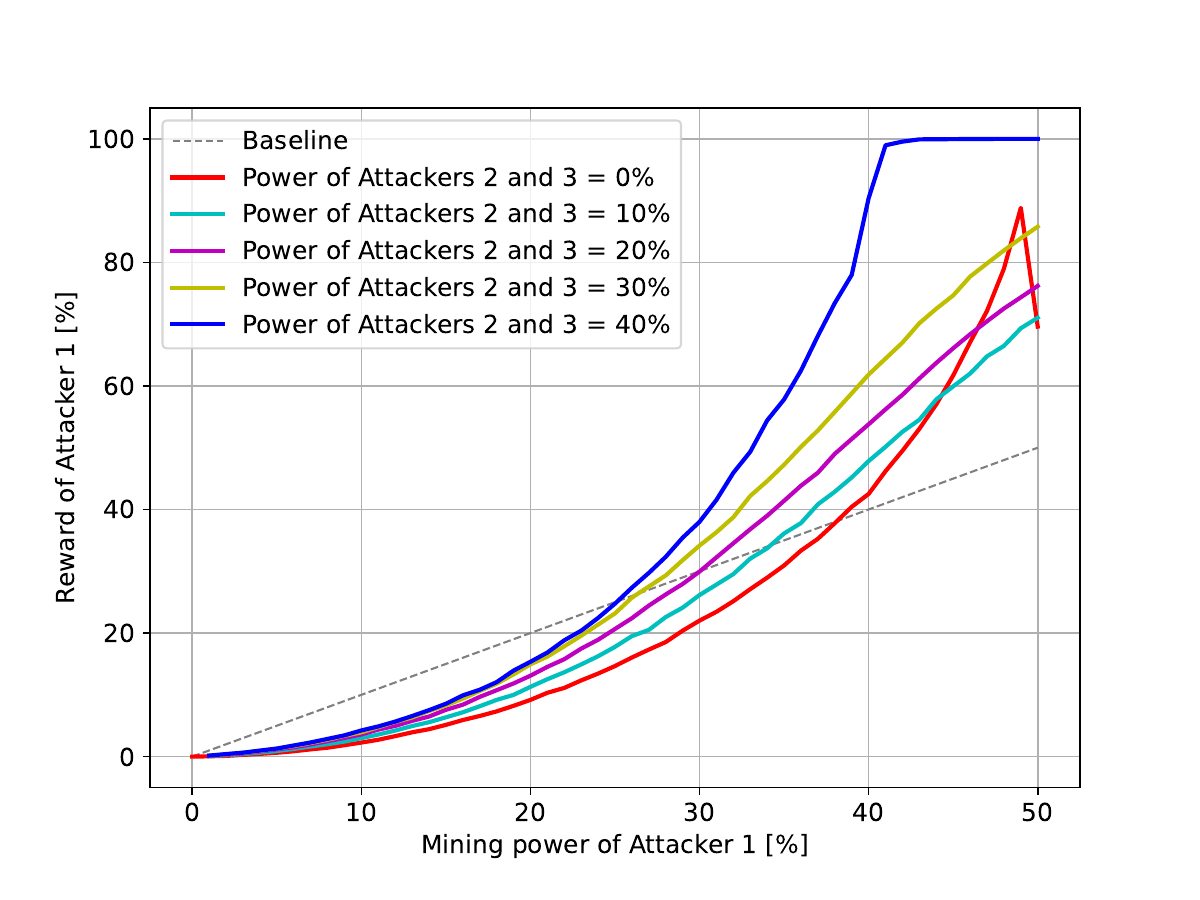}
        \caption{Longest chain rule with 3 attackers.}
        \label{fig:fruit_longest_three}
    \end{subfigure}
    \hfill
    \begin{subfigure}[b]{0.48\textwidth}
        \centering
        \includegraphics[width=\textwidth]{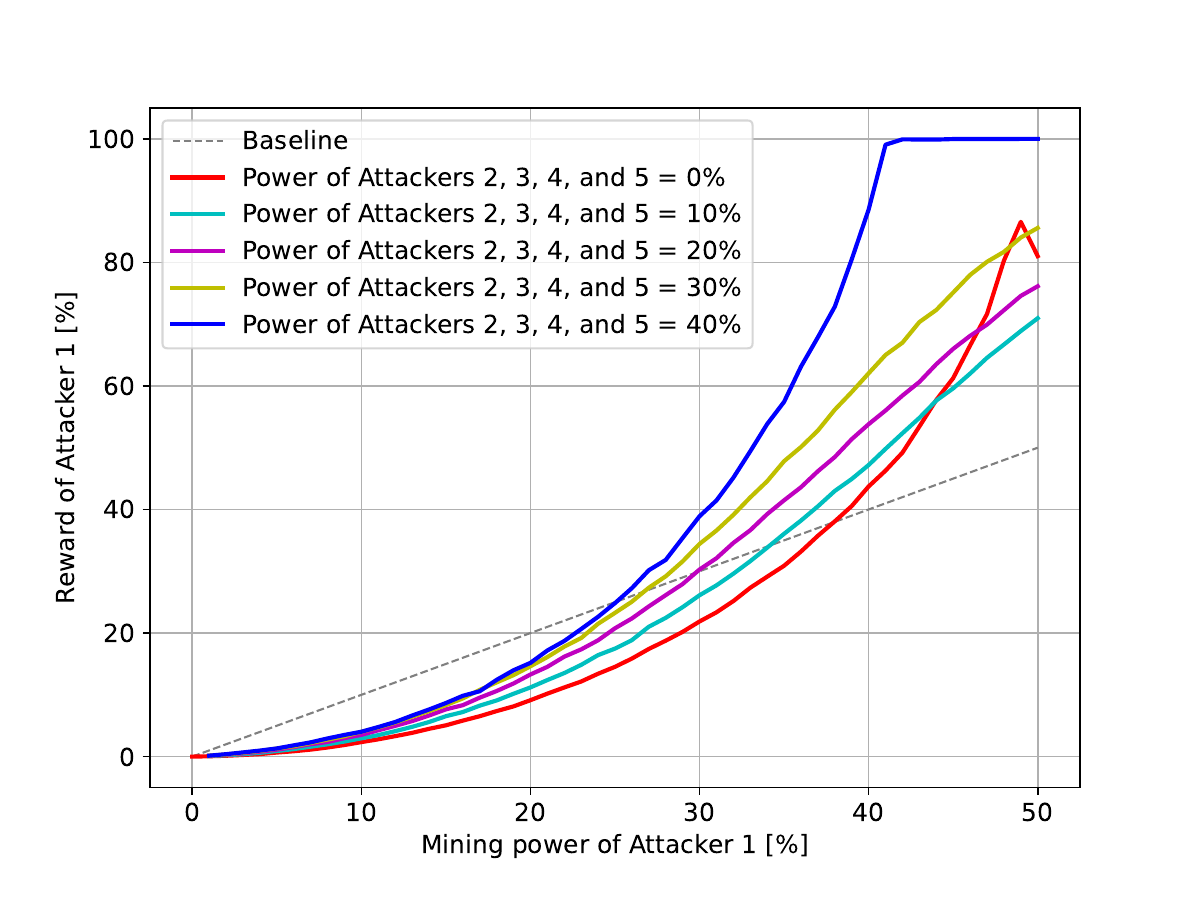}
        \caption{Longest chain rule with 5 attackers.}
        \label{fig:fruit_longest_five}
    \end{subfigure}
    \caption{Analysis of selfish mining on Fruitchain's consensus with multiple attackers. Shows the rewards of the first attacker based on their mining power relative to the other attackers with a longest chain rule for scenarios with 3 and 5 total attackers.}
    \label{fig:fruitchain_multi_attacker}
    \vspace{-1.5em}
\end{figure*}

Our experiments with Fruitchain revealed interesting dynamics, especially when considering multiple attackers.
In \autoref{fig:fruit_longest_two}, for all investigated $\gamma \in \{0,0.5,1\}$, increasing the other attacker’s power causes decrease of Attacker 1' relative revenue $R_1$, but on the other hand, it decreases his profitability threshold.
In particular, when Attacker 2 has 0\% (in red), Attacker 1 threshold is 38-39\% (i.e., as in a single attacker scenario -- see \autoref{fig:fruit_longest}).
At 20\% for Attacker 2, the profitability threshold decreases by several percentage points, and at 40\%, Attacker 1 becomes profitable at yet lower $\alpha_1$ until the point where a dominant rival can suppress his revenue (the right-hand drop).
In sum, the Attacker 1's threshold depends inversely proportional to the Attacker's 2 mining power until Attacker's 2 thresholds is reached -- meaning that Attacker 2 loses more rewards when its mining power grows since he adopts the honest chain or Attacker's 1 chain more often (while he does not know about the bigger Attacker's 1 secret chains).

\paragraph{\textbf{Three and Five Attackers.}}
This experiment considered multiple attackers, and we fixed the $\gamma$ parameter to 0.5 since it has no impact on the results.
The results are depicted in \autoref{fig:fruit_longest_three} and \autoref{fig:fruit_longest_five}, where we experimented with varying mining power of Attacker 1, while we fixed the mining power of other attackers at $\alpha_i \in \{10,20,30,40\}$\%
We can see here the same pattern as in the previous experiment (see \autoref{fig:fruit_longest_two}). 

The key observation is that a higher number of attackers benefit from the decreased profitability threshold due to multiple attackers.
However, increased relative revenue of all attackers holds only until the saturation point is reached when increasing $\alpha$ -- i.e., the sum of all relative profits cannot exceed $100\%$ / not all attackers will benefit from the attack. 

With Fruitchain the simulation ends when one of the participants' blockchains reaches the target height.
\autoref{fig:fruit_longest} shows the result of a single attacker while utilizing the longest chain rule.
We also experimented with block reward reduction from 50\% to 9.09\% (i.e., fruits are more valuable) which made the gap even larger.
This reduction also caused the point of highest negative difference to shift from baseline to higher mining power.

\section{Discussion}
The results of our empirical evaluation highlight the varying degrees of resilience that different consensus mechanisms offer against selfish mining.
Both Strongchain and Fruitchain provide a measurable improvement over the standard Nakamoto consensus, particularly in a single attacker scenario.

Strongchain consistently demonstrates the highest resistance, with a single-attacker threshold of 46\% and a two-attacker threshold of 32\%.
The core mechanism of the protocol, which incorporates weak headers in chain weight calculations, appears to be highly effective in mitigating the primary selfish mining strategy of orphaning honest blocks.
Fruitchain also shows a significant improvement, raising the single attacker threshold to 38\%.
Our findings indicate that defense capability of Fruitchain and Strongchain does not rely on network assumptions, as the $\gamma$ parameter has little effect.
A key finding across all protocols is that as the number of independent attackers increases, the individual mining power required for each to be profitable decreases.
However, the inclusive principles in Strongchain and Fruitchain ensure that the profitability thresholds remain considerably higher than in Nakamoto consensus, making collusion a more difficult and less profitable endeavor.

\section{Conclusion}
\label{conclusion}
In this work, we address the persistent threat of selfish mining with a particular focus on the under-researched area of multi-attacker scenarios.
We introduced a versatile and extensible simulation framework to conduct a rigorous comparative evaluation of selfish mining resistance across multiple Proof-of-Work consensus protocols.
By modeling Nakamoto consensus, Fruitchain, and Strongchain, we were able to validate existing findings and, more importantly, provide novel findings.

Our key findings confirm that both Strongchain and Fruitchain offer substantial improvements over Nakamoto consensus.
Strongchain is shown to be the most resilient, increasing the two-attacker profitability threshold from 21\% to 32\% per attacker.
Fruitchain also provided significant defense, raising the single-attacker threshold to 38\% and demonstrating that this security is largely independent of network connectivity advantages.
Our work provides the empirical multi-attacker thresholds for these advanced protocols, filling a gap in the literature.

This study is subject to certain limitations.
The simulation framework abstracts network dynamics into the $\gamma$ parameter, a simplification that facilitates comparison but overlooks the demonstrated impact of block propagation time and network topology on attack profitability~\cite{gervais}.
Furthermore, the model does not include transaction fees or accidental forks.
While these abstractions are standard and facilitate clear comparison, they represent areas for more detailed future analysis.
The nature of our framework paves the way for future research.
It can be extended to include other consensus protocols (e.g., Subchain~\cite{subchain}), or different consensual parameters to tweak their security even more, or other attack strategies, such as stubborn mining.
A deeper investigation into the economic incentives created by different reward structures, such as fruit and block rewards in Fruitchain, would also be a valuable contribution.
We would also like to focus on selfish mining attacks executed on DAG-based protocols (e.g. PHANTOM/GOSTDAG~\cite{phantom}, Prism~\cite{prism}) while assuming transaction fees, which have already been proven to be vulnerable to other incentive-based attacks~\cite{dagsword,arxdags}.

\paragraph{\textbf{Acknowledgment.}}
This work was supported by the internal Brno University of Technology project FIT-S-23-8151 and funded by the EU NextGenerationEU through the Recovery and Resilience Plan for Slovakia under project No. 09I05-03-V02-00057. The work was also supported by the Chips JU and its members, the DistriMuSe project, Grant Agreement No. 101139769.

\bibliographystyle{splncs04}
\bibliography{sample-bibliography}

\end{document}